\documentclass[twocolumn,showpacs,prl,aps]{revtex4-1}

\usepackage{dcolumn}
\usepackage{amssymb}

\begin{document}

\title{Short remarks on the so-called fluctuation theorems 
and related statements}

\author{Yuriy E. Kuzovlev}
\email{kuzovlev@fti.dn.ua} \affiliation{Donetsk
Physics and Technology Institute, 83114 Donetsk, Ukraine}


\begin{abstract}
It is demonstrated that the ``generalized %
fluctuation-dissipation theorem'' %
[Physica A 106, 443 (1981)] covers %
the later %
suggested ``fluctuation theorems'' %
and related statistical equalities.
\end{abstract}

\pacs{05.20.-y, 05.30.-d, 05.60.-k, 05.70.-a}

\maketitle


{\bf 1\,}\, The field of science I will talk about is the %
Hamiltonian statistical mechanics, that is statistical mechanics %
of physical systems whose all parts, - including %
thermostats (!), - can be described, %
after a suitable (canonical) choice of %
independent variables, %
 by Hamilton %
equations or equivalent Liouville equation for %
probability measures in the system's phase space %
(or, in quantum case, the von Neumann equation %
for system's density matrix). %
In this field, as far as I know, %
statistical relations like what are presently termed %
``fluctuation theorems'' (or ``work fluctuation %
theorems'', etc.) for the first time were derived %
by G.\,Bochkov and me in \cite{bk1,bk2,bk3}. %
They were named \cite{bk3}  ``nonlinear fluctuation-dissipation %
relations'' (FDR) or ``generalized fluctuation-dissipation %
theorems''.

Simultaneously in \cite{bk1,bk2,bk3,bk4,bk5,bk6,bk7,bk8,pj,d}  %
and later in %
\cite{bk9,bkt,i1,i3,i2,bk10,fdr,fdr1,dvr,dvr1} %
we demonstrated possibilities of their applications %
to

$\ast$\,\, construction of such Marcovian models of %
noise, relaxation, transport, and fluctuations in %
non-equilibrium steady states, what %
agree with all the FDR and consequently are  %
thermodynamically correct \cite{bk1,bk2,bk3,bk4,bk5,bk6,d};

$\ast$\,\, canonic construction of kinetic potentials %
and variational principles  %
for non-equilibrium thermodynamics \cite{bk4,bk8,pj};

$\ast$\,\, derivation of various particular relations %
between non-linear responses, fluctuations of responses %
and higher-order statistical %
characteristics of %
noise \cite{bk1,bk2,bk7,bkt,d} (for example, between %
nonlinear optical  susceptibilities and fluctuations of Raman %
scattering in transparent media \cite{d});

$\ast$\,\, analysis of phase volume exchange and non-linear %
reciprocity relations between different channels of %
transport and dissipation \cite{bkt};

$\ast$\,\, revealing connections between statistics %
of transport processes (e.g. counting statistics of %
charge transfer) and a shape of dissipative %
non-linearity (e.g. CVC) \cite{bk2,bk3,bk8,bkt,fdr1,d};

$\ast$\,\, description of continuous %
quantum measurements and construction of formally %
exact stochastic evolution (Liouville) equations %
for open systems \cite{fdr,fdr1};

$\ast$\,\, construction of formally exact Langevin %
equations for open systems \cite{fdr1};

$\ast$\,\,  analysis of fluctuations in work,  dissipation, %
relaxation and transport rates, and  %
entropy production,  in systems driven by %
dynamical forces (entering the system's Hamiltonian) %
or/and thermic forces (entering system's initial %
probability distribution) %
\cite{bk1,bk2,bk3,bk9,i1,i3,i2};

$\ast$\,\, revealing and description of fundamental %
scaleless low-frequency %
(1/f\,-type) fluctuations in rates of irreversible %
processes, in particular, in mobilities and %
diffusivities %
of charge carriers and fluid molecules %
\cite{bk9,i1,i3,i2,bk10,dvr,dvr1,f1}.

\,\,\,

Some of the above references may supplement %
the G.\,Crooks' \cite{cr} %
and later \cite{rmp,h} bibliographies on %
the ``fluctuation theorems'' and related %
statements. %

A part of the  works enumerated in these %
bibliographies deals with not only %
Hamiltonian dynamics but also with systems %
represented by %
Marcovian random processes, as in \cite{cr1}, %
may be, degenerating into %
deterministic %
dissipative processes directed by  %
non-physical toy %
(zero-temperature, phase volume eating) %
thermostats, as in \cite{e}. %
Validity of ``fluctuation theorems'' in such %
model systems is not %
surprising, because of the %
above mentioned reproducibility of FDR in %
properly constructed Marcovian  %
models \cite{f2}.

The really new things, %
in comparison with \cite{bk1,bk2,bk3}, %
are, for the first look, such relations as the %
``Crooks fluctuation theorem'' \cite{cr,rmp,h,cr1} %
and the ``Jarzynski equality''  \cite{rmp,h,jar} %
which visually are significantly different from %
similar relations of \cite{bk1,bk2,bk3}. %
Sometimes one can read that the latter are %
``more limited'' (or ``less general'', etc.) %
than the former. However, this is wrong %
opinion (perhaps, caused by laziness of thinking), %
and below I will demonstrate its falseness.

\,\,\,

{\bf 2}.\,\, Let us consider a Hamiltonian system %
disturbed by some external forces $\,x=x(t)\,$ %
entering its Hamiltonian $\,H=H(x(t),\Gamma)\,$, %
where $\,\Gamma\,$ denotes a full set of (canonic) %
variables of the system. Anyway, a character %
of reaction of the system to constant forces, %
$\,x=\,$const\,, is of principal importance. %
In one case, sufficiently weak constant forces %
lead to proportionally weak stable changes in %
the system's state. In the opposite case, %
arbitrary weak constant forces can induce %
with time some finite or even arbitrary strong changes %
(the examples are responses of an isolator and a %
conductor to external electric field). %

In \cite{bk1,bk2,bk3}, the second case was %
in the centre of attention. At that, clearly, %
it is reasonable to write
\begin{equation}
\begin{array}{c}
H(x,\Gamma)\,=\, H_0(\Gamma)-h(x,\Gamma)\,\,\,,\\
h(x,\Gamma)\,=\, x\cdot Q(\Gamma)\,\,\,, \label{h0}
\end{array}
\end{equation}
with quite unambiguous division of the Hamiltoniam %
into main unperturbed part, $\,H_0\,$, and its %
perturbation, $\,h\,$. %
But the first case also was foreseen in \cite{bk1} %
and \cite{bk3}. At that, however, separation of main %
and perturbed parts of Hamiltonian  no longer is %
unambiguous. Therefore we can write
\begin{equation}
\begin{array}{c}
H(x,\Gamma)\,=\, H_0(\Gamma)-h(x,\Gamma)\,\,\,,\\
H_0(\Gamma)\,=\, H(x_0,\Gamma)\,\,\,,\\
h(x,\Gamma)\,=\, H(x_0,\Gamma)- %
H(x,\Gamma)\,\,\,, \label{h1}
\end{array}
\end{equation}
with some formally arbitrary $\,x_0\,$.

Since, evidently, both the %
``Crooks fluctuation theorem''  %
and ``Jarzynski equality''  %
presume just such case, we will exploit the %
arbitrariness of $\,x_0\,$. %
Concretely, considering change of the external %
forces from $\,x(0)\,$ to $\,x(\theta)\,$ %
during time interval $\,0<t<\theta\,$, %
let us choose\, %
\begin{equation}
\begin{array}{c}
x_0\,=\,x(\theta)\,\,\,, \nonumber
\end{array}
\end{equation}
and then choose the initial probability %
distribution to be, as in \cite{bk1}, %
\begin{eqnarray}
\rho_0(\Gamma)\,=\,  \rho(x_0,\Gamma)\,\equiv\, %
\exp{\{\beta\,[F(x_0) %
-H(x_0,\Gamma)\,]\}}\,\,\,,\nonumber
\end{eqnarray}
where
\begin{eqnarray}
\exp{\{-\beta\,F(x)\}}\,\equiv\, %
\int \exp{\{ -\beta\,H(x,\Gamma)\}}\,d\Gamma %
\,\,\,\nonumber
\end{eqnarray}
This is equilibrium distribution formed %
before $\,t=0\,$ under constant forces $\,x_0\,$. %
It, however, %
becomes non-equilibrium after instant jump %
of the external forces from $\,x_0\,$ to $\,x(0)\,$. %
According to \cite{bk1,bk3}, %
the exact identity takes place, %
\begin{eqnarray}
\langle\, \exp{(-\beta\,E)}\, \rangle\, \equiv\, %
\nonumber\\ \,\equiv\, %
\int \exp{[\,-\beta\,E(\theta,\Gamma)]}\, %
\rho(x_0,\Gamma)\, d\Gamma\, %
 \,=\, 1\,\,\,,\label{bk}
\end{eqnarray}
where \cite{bk1}
\begin{eqnarray}
E\,=\,E(\theta,\Gamma) \,\equiv\, %
H_0(\Gamma(\theta))\,-\,H_0(\Gamma)\,=\, \label{e}\\
\,=\, \int_0^\theta \left[\frac {d}{dt} - %
\frac {dx(t)}{dt} \cdot \frac {\partial}{\partial x(t)} %
\right]\,h(x(t),\Gamma(t))\, dt\,=\,\nonumber\\
\,=\,  \int_0^\theta %
\frac {d\Gamma(t)}{dt} \cdot %
\frac {\partial}{\partial \Gamma(t)} %
\,\,h(x(t),\Gamma(t))\, dt
  \,\,\,, \nonumber
\end{eqnarray}
with\, $\,\Gamma(t)\,$\, representing the system's %
state at time $\,t\,$ considered as the function %
of initial state $\,\Gamma\,$ %
(and, of course, a functional %
of the forces $\,x(t)\,$). %

Formally, this is trivial consequence of %
the phase volume conservation. %
Physically, since both the initial state %
at $\,t=0\,$ and final state at $\,t=\theta\,$ %
correspond to the same values $\,x_0\,$, %
we can treat the latter as reference point for %
the forces and interpret (\ref{e}) as the energy %
dissipated by the system during its %
perturbation  $\,x(t)-x_0\,$.

Now, let us rewrite the expression in %
the exponent in (\ref{bk}) as follows:
\begin{eqnarray}
F(x_0)\,-\,E(\theta,\Gamma)\,-\,H_0(\Gamma) %
\,=\, \nonumber\\ %
\,=\, F(x_0)\,-\,[\,H(x(\theta),\Gamma(\theta)) %
\,-\, H(x(0),\Gamma)\,]\,-\, H(x(0),\Gamma) %
\,\,\,,\nonumber
\end{eqnarray}
and the Eq.\ref{bk}, correspondingly, in the form %
\begin{eqnarray}
\exp{\{\beta[\,F(x(\theta))\,-\,F(x(0))\,]\}}\, %
\times\nonumber \\
\times\, %
\int \exp{[\,-\beta\,\mathcal{E}(\theta,\Gamma)]}\, %
\rho(x(0),\Gamma)\, d\Gamma \,=\, \nonumber\\
\,=\, \exp{\{\beta[\,F(x(\theta))\,-\,F(x(0))\,]\}}\, %
\times\nonumber\\ \,\times\, %
\langle\,\exp{[\,-\beta\,\mathcal{E}]}\,\rangle\,=\,1 %
\,\,\,,\label{bk1}
\end{eqnarray}
where
\begin{eqnarray}
\mathcal{E}\,=\,\mathcal{E}(\theta,\Gamma)\, %
\equiv\, H(x(\theta),\Gamma(\theta))\,-\, %
H(x(0),\Gamma)\,=\nonumber\\
=\,\int_0^\theta \frac {dx(t)}{dt}\cdot %
\frac {\partial H(x(t),\Gamma(t))} %
{\partial x(t)}\, dt\,\, \label{e1}
\end{eqnarray}
is full change of energy of the system, - %
consisting of changes in internal energy %
(i.e. dissipation) and energy of interaction %
with external forces, - %
under the same time variation of the forces %
as in Eq.\ref{bk} \cite{f3}. %
Obviously, Eq.\ref{bk1} is nothing but %
the ``Jarzynski equality'' \cite{jar}. %
Thus, in essence, it is particular consequence %
of the equalities obtained in \cite{bk1} %
or \cite{bk3}.

In quite similar way one %
can show that general relations for %
probability and characteristic functionals %
of phase trajectories, %
found in \cite{bk1} and \cite{bk3}, %
imply \cite{f4} the Crooks' and other %
``fluctuation theorems'' \cite{cr,rmp,h,cr1}. %
For instance, relations like (7) from %
\cite{bk1} or (1) from \cite{bk2} %
in the case (\ref{h1})\,  %
easy transform into %
\begin{eqnarray}
P(\widetilde{\mathcal{T}};\widetilde{x})/ %
P(\mathcal{T};x) %
\,=\, \exp{\{\beta\,[\Delta F %
-\mathcal{E}]\}}\,\,\,, \label{p1}
\end{eqnarray}
where $\,\mathcal{T}\,$  means %
(more or less detalized) phase trajectory, %
the tilda means time reversion %
(e.g. $\,\widetilde{x}(t)=\epsilon x(\theta -t)\,$), %
\,$\,\Delta F\,=\, %
F(x(\theta))-F(x(0))\,$\,, %
and statistical ensemble %
is the same as in (\ref{bk1}). %
Consequently, %
\begin{eqnarray}
P(-\mathcal{E};\widetilde{x})/ %
P(\mathcal{E};x) %
\,=\, \exp{\{\beta\,[\Delta F %
-\mathcal{E}]\}}\,\,\,. \nonumber
\end{eqnarray}

{\bf 3}.\,\, In conclusion, I should notice %
that quantum version of the transition from %
Eq.\ref{bk} to Eq.\ref{bk1} can be formulated %
by exact analogy with \cite{bk1,bk3}. %
Ar the same time, I should admit that quantum %
FDR still are investigated very insufficiently, %
and therefore the present activity in the field %
of ``quantum fluctuation theorems'' \cite{h} (see %
also bibliography in \cite{cr,rmp,h}) %
undoubtedly might produce %
many original results.



\begin{thebibliography}{27}
\bibitem{f1}
Usefulness of FDR %
in this respect was confirmed by comparison with direct %
analysis of the Liouville equation \cite{tmf}.
\bibitem{f2}
The basic FDR are insensitive to internal structure %
of a system under consideration, %
in particular, to structure of its internal thermostats.
\bibitem{f3}
Of course, the angle brackets in Eq.\ref{bk} %
and Eq.\ref{bk1} are not one and the same %
operation, because symbolize statistical averages %
over different initial ensembles. %
\bibitem{f4}
Notice that formally the variables %
$\,Q=Q(\Gamma)\,$ ($\,\Gamma =\{\alpha,\beta\}\,$), %
$\,I=I(\Gamma)\,$, etc., in \cite{bk1,bk2,bk3} %
can include all the set $\,\Gamma\,$ and %
besides arbitrary sets of functions of $\,\Gamma\,$ %
(at that, of course, any of the conjugated forces %
may be in fact time-independent, e.g. %
merely equal to zero).
\end{thebibliography}



\end{document}